\documentclass[10pt,aps,prc,floatfix,twocolumn,nofootinbib,superscriptaddress]{revtex4-1}
\usepackage{graphicx,amsmath,amssymb,bm}
\usepackage{amsfonts}
\usepackage[utf8]{inputenc}
\usepackage{verbatim}
\usepackage{float}
\usepackage[labelfont={small},font=small,subrefformat=parens,caption=false]{subfig}
\usepackage{cancel}
\usepackage{multirow}
\usepackage{array}
\usepackage{xparse}
\usepackage{xspace}

\usepackage{bigstrut}

\usepackage{braket}
\usepackage[ISO]{diffcoeff}

\usepackage{color}
\usepackage{dsfont}
\usepackage{dcolumn}

\usepackage[pdfencoding=auto, pdfpagelabels]{hyperref}

\usepackage[overload]{textcase}

\newcommand*{\diffdchar}{\mathrm{d}}    % or {ⅆ}, or {\mathrm{d}}, or whatever standard you’d like to adhere to
\newcommand*{\dd}[1]{\mathop{\diffdchar #1}}

\newcommand{\action}{\mathcal{S}}
\newcommand{\nbasis}{N_b}
\newcommand{\ntrain}{N_\mathrm{sample}}

\graphicspath{{figures/}{template/}}

\newcommand{\trial}{\widetilde}
\newcommand{\subspace}{\widetilde}
\newcommand{\coeff}{\beta}
\newcommand{\coeffs}{\vec{\coeff}}

\newcommand{\params}{\boldsymbol{\theta}}

\newcommand{\eg}{\textit{e.g.}\xspace}
\newcommand{\ie}{\textit{i.e.}}
\newcommand{\etal}{\textit{et~al.}\xspace}
\newcommand{\CC}{C\nolinebreak\hspace{-.05em}\raisebox{.4ex}{\tiny\bf +}\nolinebreak\hspace{-.10em}\raisebox{.4ex}{\tiny\bf +}}
\definecolor{linkcolor}{rgb}{0,0,0.40} 
\hypersetup{%
    pdfsubject=Paper,
    pdfkeywords={nuclear physics} {variational} {Ritz} {Galerkin} {emulator},
    unicode = true,
    breaklinks = true,
    colorlinks = true,
    linkcolor = linkcolor,
    citecolor = linkcolor,
    menucolor = linkcolor,
    urlcolor = linkcolor
}

\usepackage{cellspace}
\graphicspath{{./figures/}}

\begin{document}

\title{Model reduction methods for nuclear emulators}

\author{J.~A. Melendez}
\email{melendez.27@osu.edu}
\affiliation{Department of Physics, The Ohio State University, Columbus, OH 43210, USA}

\author{C. Drischler}
\email{drischler@frib.msu.edu}
\affiliation{Facility for Rare Isotope Beams, Michigan State University, MI 48824, USA}

\author{R.~J. Furnstahl}
\email{furnstahl.1@osu.edu}
\affiliation{Department of Physics, The Ohio State University, Columbus, OH 43210, USA}

\author{A.~J. Garcia}
\email{garcia.823@osu.edu}
\affiliation{Department of Physics, The Ohio State University, Columbus, OH
43210, USA}

\author{Xilin Zhang}
\email{zhangx@frib.msu.edu}
\affiliation{Facility for Rare Isotope Beams, Michigan State University, MI 48824, USA}

\newcommand{\aka}{also known as\xspace}

\date{\today}

%%%%%%%%%%%%%%%%%%%%%%%%%%%%%%%%%%%%%%%%%%%%%%%%%%%%%%%%
\begin{abstract} % do not use macros in the abstract

The field of model order reduction (MOR) is growing in importance due to its ability to extract the key insights from complex simulations while discarding computationally burdensome and superfluous information.
We provide an overview of MOR methods for the creation of fast \& accurate emulators of memory- and compute-intensive nuclear systems.
As an example, we describe how ``eigenvector continuation'' is a special case of a much more general and well-studied MOR formalism for parameterized systems.
We continue with an introduction
to the Ritz and Galerkin projection methods that underpin many such emulators, while pointing to the relevant MOR theory and its successful applications along the way.
We believe that this will open the door to broader applications in nuclear physics and facilitate communication with practitioners in other fields.

\end{abstract}
%%%%%%%%%%%%%%%%%%%%%%%%%%%%%%%%%%%%%%%%%%%%%%%%%%%%%%%%

\maketitle

\section{Introduction}
\label{sec:introduction}

Nuclear physics calculations often need to be repeated many times for different values of some model parameters,
for example when sampling the model space for Bayesian uncertainty quantification~\cite{Neufcourt:2019qvd,King:2019sax,Ekstrom:2019lss,Catacora-Rios:2020xgx,Wesolowski:2021cni,Svensson:2021lzs,Odell:2021tqd,Djarv:2021pjc,Alnamlah:2022eae} and experimental design~\cite{Melendez:2020ikd,Phillips:2020dmw,Farr:2021fyc}.
The computational burden can be alleviated by using \emph{emulators}, or surrogate models, which accurately approximate the response of the 
original (\ie, high-fidelity)
model but are much cheaper to evaluate.
The development of emulators by nuclear physicists is a welcome addition to the community's toolkit, but such applications are neither the first nor the most sophisticated in the long history of \emph{model reduction} for complex simulations.

The na\"ive implementation of realistic computer models in physics, mathematics, or engineering generally can demand an ever-increasing computational burden, but it has been known for decades that much of the information contributing to this burden is superfluous and can be \emph{compressed} into a much more efficient form while retaining high accuracy.
The broad and relatively mature field of model order reduction (MOR) has focused on exactly this problem of extracting the dominant information while excluding the costly, redundant information from simulations; the resulting emulator is known as the \emph{reduced-order model}.
Initially, the systems under consideration contained only fixed sets of model parameters, but later on the study of parametric MOR (PMOR) addressed the need to emulate systems as the parameter values changed.

The problems already addressed by PMOR are wide-ranging; 
for example, a sampling of applications 
in engineering, 
computational physics, and computer science is presented in Ref.~\cite{Benner2020Volume3Applications}.
The problems include large-scale systems of ordinary and 
partial differential equations with time dependence and nonlinearities, eigenvalue problems, and much more.
Although these particular examples are classical, the nuclear physicist is likely to find problems in the MOR literature with analogous mathematical structure and goals to their own problems. 
In later sections we point to examples in the quantum realm such as bound-state and scattering problems and density functional theory.

Reduction schemes in the MOR literature can be classified as data-driven or model-driven, although hybrid approaches are also possible.
Data-driven approaches typically need little to no understanding of the system under consideration, but rather rely on interpolating the output of the high-fidelity model.
These are hence classified as \emph{non-intrusive}.
Examples include Gaussian processes~\cite{rasmussen2006gaussian} and dynamic mode decomposition (DMD)~\cite{doi:10.1146/annurev-fluid-011212-140652,KutzDMDbook2016}, and there are further exciting developments along these lines coming from the machine learning literature (\eg, Refs.~\cite{raissi2019physics,CHEN2021110666,FRESCA2022114181}).

Alternatively, the model-driven approaches take the high-fidelity system of equations as given, from which they derive the reduced-order equations.
Emulation via model-driven methods continue to be physics-based, respecting the underlying structure of the system, and are likely to extrapolate more effectively than data-driven emulators.
Many model-based methods employ the concept of projection, where the high-dimensional system is projected onto a well-chosen low-dimensional manifold.
The challenge of many model-driven approaches is related to their \emph{intrusive} nature, \ie, they require writing new codes for projecting the system of interest into a reduced space.

Despite the mature literature on MOR, many of its aspects have been reinvented across multiple different disciplines.
Indeed, the nuclear physics community has recently been exploiting an already-established model-driven, projection-based PMOR technique in the study of the eigenvalue problem.
Introduced to the nuclear community as eigenvector continuation (EC)~\cite{Frame:2017fah,Sarkar:2020mad}, this method has been demonstrated to be highly effective for nuclear bound-state~\cite{Konig:2019adq, Ekstrom:2019lss, Wesolowski:2021cni,Yoshida:2021jbl} and scattering calculations~\cite{Furnstahl:2020abp,Melendez:2021lyq,Drischler:2021qoy, Zhang:2021jmi}.
EC has not been recognized in its broader context because the projection techniques common in the model reduction literature are not widely known in the nuclear physics community.
Furthermore, many reviews of the subject 
presume a level of maturity with these tools that might make them difficult to digest.

We therefore provide a guide to projection-based PMOR for the nuclear physics community.
Our goal in the present work is not to introduce specific new nuclear physics applications, but to make the existing literature more accessible.
In particular, we explain how one can obtain emulators by projecting generic differential equations and eigenvalue problems into a low-dimensional subspace, all while relating key ideas back to the broader MOR literature.
By doing so, we seek to open the door to broader applications, allow the nuclear community to take advantage of the vast literature, and facilitate communication with practitioners in other fields.

In Sec.~\ref{sec:setting_the_stage} we begin by introducing important concepts and notation.
We then provide an intuitive example of PMOR as applied to the eigenvalue problem in Sec.~\ref{sec:eigen} and place existing work in the nuclear physics literature into a broader context.
Next, Secs.~\ref{sec:variational} and~\ref{sec:galerkin} provide two distinct methods for projecting differential equations onto effective low-dimensional manifolds via variational principles (or, the Ritz method) and the Galerkin projection, respectively.
Finally, we return in Sec.~\ref{sec:model_reduction} to the recent developments made in the MOR community to discuss the incredible opportunities available to nuclear physicists interested in applying such techniques.
An outlook is given in Sec.~\ref{sec:outlook}.

\section{Setting the stage}
\label{sec:setting_the_stage}

We would like to solve a differential equation or an eigenvalue problem where the operators are a function of parameters $\params$.\footnote{
Note that the literature may use a different notation, including $\mu$ for the set of parameters.}
In the case of a differential equation, the goal is to obtain the solution $\psi$ of
\begin{subequations} \label{eq:generic_differential_and_boundary}
\begin{align}
    D(\psi; \params) & = 0 \quad \text{in } \Omega  , \label{eq:generic_differential} \\
    B(\psi; \params) & = 0 \quad \text{on } \Gamma , \label{eq:generic_boundary}
\end{align}
\end{subequations}
where $\{D, B\}$ are operators, and $\{\Omega, \Gamma\}$ are the domain and boundary, respectively.
Generalizations to systems of differential equations follows straightforwardly.
For the eigenvalue problem the solutions are $\{E, \ket{\psi} \}$, which satisfy
\begin{align} \label{eq:generic_eigenvalue_problem}
    H(\params)\ket{\psi} = E\ket{\psi}
\end{align}
for a given Hermitian operator $H$.
Throughout this work we switch between an abstract vector notation $\ket{\psi}$ and functions $\psi$ with the representation dependencies suppressed.
Time-dependence is permitted in these systems but is not explicitly considered here---see Sec.~\ref{sec:model_reduction} for pointers on how to handle these cases.

Here we consider systems where obtaining $\ket{\psi}$ (and $E$) will require a non-negligible amount of computing time, which becomes compounded when a range of parameter values are required, \eg, in a Monte Carlo sampler or an optimizer.
The choice in PMOR is then made to spend compute resources in the \emph{offline stage}, where the heavy lifting can be easily parallelized in many cases~\cite{benner2015survey}, such that emulation during the \emph{online phase} can be performed efficiently.
A critical component of the online-offline paradigm is the removal of all size-$\psi$ operations during the online phase, a property available for operators with an affine parameter dependence.
That is,
\begin{align} \label{eq:H_affine}
    H(\params) = \sum_n h_n(\params) H_n,
\end{align}
or likewise for $D(\psi; \params)$ or $B(\psi;\params)$, where the operators can be written as a sum of products of parameter-dependent functions $h_n(\params)$ and parameter-independent operators $H_n$.
This factorization permits the operators to be projected \emph{once} up front, rather than for every value of $\params$ we would like to emulate, and will be discussed in more detail in the following sections.
For non-linear systems and cases with non-affine parameters, various \emph{hyperreduction} methods have been developed to augment the model reduction techniques (see Secs.~\ref{sec:variational_results_nuclear} and~\ref{sec:Building_the_ROM}).

The projection-based emulation approaches described here rely on (i) choosing an effective low-dimensional representation of $\psi$ and (ii) writing Eqs.~\eqref{eq:generic_differential_and_boundary} and~\eqref{eq:generic_eigenvalue_problem} in integral form.
For the first step one proposes a trial function comprised as a linear combination of a set of $\nbasis$ known basis functions $\{\psi_i\}$
\begin{align} \label{eq:trial_general}
    \trial\psi & \equiv \sum_{i=1}^{\nbasis} \coeff_i\psi_i = X\coeffs , \\
    X & \equiv
    \begin{bmatrix}
        \psi_1 & \psi_2 & \cdots & \psi_{\nbasis}
    \end{bmatrix}.
\end{align}
While the coefficients $\coeffs$ are unspecified at this point, their values will become fixed after imposing conditions specific to the integral forms described below.
We focus on bases $X$ that are constructed out of \emph{snapshots}~\cite{Benner2017modelRedApprox,Benner20201,Buchan2013EigenvaluePOD,Quarteroni:218966}, high-fidelity solutions $\psi_i = \psi(\params_i)$, at a set of parameter values $\{\params_i\}$.
This requires that a high-fidelity solver for $\psi$ exists, though the details of such a solver are irrelevant for our discussion.
The snapshot approach has been found to construct highly efficient and system-specific trial bases across a wide range of cases.
We will return to exactly how these snapshots can be chosen in Sec.~\ref{sec:model_reduction}.
This form of the trial function~\eqref{eq:trial_general} will be used throughout the rest of this work.
Colloquially, we will describe the space of $\nbasis$ basis functions as the ``small space'' and the space of $\psi$ and its corresponding operators ($D$, $B$, and $H$) as the ``large space.''

One familiar integral form available to many differential equations comes from variational principles.
Variational principles begin with the definition of a scalar functional $\action$ (the action) that can be written as
\begin{align} \label{eq:action}
    \action[\psi] = \int_\Omega \dd{\Omega} F(\psi) + \int_\Gamma \dd{\Gamma} G(\psi),
\end{align}
where $F$ and $G$ are known differential operators.
The unknown function $\psi$ is determined as the one that makes $\action$ \emph{stationary}, \ie, $\delta\action = 0$, under arbitrary variations $\delta\psi$.
This will form the basis of variational emulators discussed in Sec.~\ref{sec:variational}.

It is not always the case that Eq.~\eqref{eq:generic_differential_and_boundary} can be cast in the form of a variational principle as in Eq.~\eqref{eq:action}.
In this case we can instead turn to the \emph{weak form} of Eq.~\eqref{eq:generic_differential_and_boundary} whose errors or residuals we aim to minimize~\cite{zienkiewicz2013finite}.
This will form the basis of the more general Galerkin emulators discussed in Sec.~\ref{sec:galerkin}.

However, before exploring the more abstract variational and Galerkin approaches, we begin in Sec.~\ref{sec:eigen} with a concrete and (to physicists) familiar example of both: the eigenvalue problem.
This contains many of the ideas fundamental to these more general cases but is likely a more accessible place to start.

\section{Eigen-Emulators}
\label{sec:eigen}

\subsection{Derivation}

The eigenvalue problem frequently appears in nuclear physics when solving the few- or many-body Schr{\"o}dinger equation.
We are often particularly interested in the ground state with energy $E_{\text{min}}$ and the associated wave function, but the full spectrum of eigen-energies and wave functions may also be of interest.
More generally, eigenvalues appear across the physical sciences and engineering due to their appearance in the solution of differential equations, and come with a broad range of associated boundary conditions.
Here we focus on converting the simple case to a form amenable to reduced-order modeling, and we point to the literature on variational forms of eigenvalue problems for more specific cases~\cite{BABUSKA1991641}.

We can write down the following functional whose stationary solution approximates the eigen-energy, that is $\mathcal{S} = \mathcal{E}$, where
\begin{align} \label{eq:eigenvalue_variational_principle}
    \mathcal{E}[\psi] & = \braket{\psi | H | \psi} - \subspace{E} (\braket{\psi | \psi} - 1)
\end{align}
and the normalization of the wave function has been imposed with a Lagrange multiplier $\subspace{E}$.
Under arbitrary variations $\delta\psi$, the change in $\mathcal{E}$ for Hermitian $H$  can be written as
\begin{align}
    \delta\mathcal{E}[\psi] & = 2\braket{\delta\psi | [H-\subspace{E}] | \psi} - \delta\subspace{E} ({\braket{\psi | \psi}} - 1).
\end{align}
Assuming $\ket{\trial\psi} = X\coeffs$ as in Eq.~\eqref{eq:trial_general}, we then find that the values of $\coeffs$ satisfying the stationary condition $\delta\mathcal{E}[\trial\psi_\star] \equiv 0$, denoted $\coeffs_\star$, are those obtained from the following generalized eigenvalue problem
\begin{align}
    \subspace{H}\coeffs_\star & = \subspace{E}\subspace{N} \coeffs_\star ,\label{eq:generalized_eigenvalue} \\
    \coeffs_\star^\dagger \subspace{N} \coeffs_\star & = 1 ,\label{eq:generalized_eigenvalue_normalization}
\end{align}
where $\subspace{H} \equiv X^\dagger H X$ is the Hamiltonian projected into the subspace spanned by $X$, and $\subspace{N} \equiv X^\dagger X$ is the norm matrix.
The meaning of $\subspace{E}$ can be understood by substituting these relationships back into the variational form~\eqref{eq:eigenvalue_variational_principle}:
\begin{align}
    \mathcal{E}[\trial\psi_\star] = \subspace{E} \coeffs_\star^\dagger \subspace N \coeffs_\star = \subspace{E}.
\end{align}
Thus $\subspace{E}$ is an approximation to the energy.
Equations~\eqref{eq:generalized_eigenvalue} and~\eqref{eq:generalized_eigenvalue_normalization}, combined with Eq.~\eqref{eq:trial_general}, therefore constitute the reduced-order model for $E$ and $\ket{\psi}$ projected to the $\nbasis \times \nbasis$ small space.
In the context of the Rayleigh-Ritz method%
\footnote{%
See Refs.~\cite{LEISSA2005961,ILANKO2009731} for commentary on the history of the method name.%
} 
$\subspace{E}$ is known as the \emph{Ritz value} while $\trial\psi_\star = X\coeffs_\star$ is known as the \emph{Ritz vector}.

We have seen how a variational principle for an eigenvalue problem can lead to a reduced-order model for both $E$ and $\psi$.
But there is a more general approach to arrive at exactly Eqs.~\eqref{eq:generalized_eigenvalue} and~\eqref{eq:generalized_eigenvalue_normalization} that is often described as the Galerkin method.
The Galerkin approach begins with the \emph{weak} form of the eigenvalue problem~\eqref{eq:generic_eigenvalue_problem}:
\begin{align} \label{eq:eigen_value_weak_full}
    \braket{\phi | H - E | \psi} = 0,
\end{align}
where $\phi$ is an arbitrary \emph{test} function.
It can be shown that if Eq.~\eqref{eq:eigen_value_weak_full} holds for all $\phi$, then Eq.~\eqref{eq:generic_eigenvalue_problem} must hold as well: if $(H - E)\ket{\psi}$ has any non-zero elements, then one can immediately find a $\phi$ such that Eq.~\eqref{eq:eigen_value_weak_full} does not hold.

To obtain a reduced-order model from Eq.~\eqref{eq:eigen_value_weak_full} we would ideally like to map $\psi \to \trial\psi$ and find the $\coeffs$ that satisfy Eq.~\eqref{eq:eigen_value_weak_full} for all $\phi$, hence ensuring the satisfaction of Eq.~\eqref{eq:generic_eigenvalue_problem}.
Unfortunately this would result in an over-determined system because an arbitrary $\phi$ has many more degrees of freedom than $\trial\psi$.
Alternatively, we can derive the Ritz subspace method by imposing that the error made by the trial eigenvector is orthogonal to the $\nbasis$-dimensional subspace $\mathcal{X}$ spanned by $X$:
\begin{align}
    H\ket{\trial\psi} - \subspace{E}\ket{\trial\psi} &\perp \mathcal{X},
\intertext{or likewise}
    \braket{\phi | H - \subspace{E} |\trial\psi} &= 0, ~~\forall \phi \in \mathcal{X}. \label{eq:eigenvalue_galerkin}
\end{align}
Note here that, due to a peculiarity of the eigenvalue problem, we had to make the replacement $E \to\subspace{E}$ because the eigenvalue is an \emph{output} of the system: if the span of $\mathcal{X}$ does not exactly contain $\psi(\params)$ then $\subspace{E}\neq E$ in general.
Equation~\eqref{eq:eigenvalue_galerkin} is known as the Galerkin condition, and is equivalent to imposing that $\braket{\psi_i | H - \subspace{E} |\trial\psi} = 0$ must hold for $i \in [1, \nbasis]$.
This yields a system of $\nbasis$ equations and $\nbasis$ unknowns $\coeffs$ and, together with the normalization condition, reduces exactly to Eqs.~\eqref{eq:generalized_eigenvalue} and~\eqref{eq:generalized_eigenvalue_normalization}.
The Lagrange multiplier was not strictly necessary in this implementation of the Galerkin method because the normalization is irrelevant for Eq.~\eqref{eq:eigenvalue_galerkin}, but see Sec.~\ref{sec:galerkin} for a discussion on including constraints for other Galerkin problems.

The applications of the eigensystem reduced-order models to nuclear physics have benefited from
the fact that Hamiltonians derived from chiral effective field theory (EFT) have the form of Eq.~\eqref{eq:H_affine} due to their affine dependence on the parameters $\params$ called low-energy couplings.
By projecting each $H_n \to \subspace{H}_n \equiv X^\dagger H_n X$ in the offline stage, $\subspace{H}(\params)$ can be efficiently reconstructed for each new $\params$, and Eq.~\eqref{eq:generalized_eigenvalue} rapidly solved, in the online stage.
Furthermore, if the $h_n(\params)$ are \emph{smooth}, as they are for chiral EFT, then we have found that the low-dimensional representations exploited by the reduced-order models are often quite well satisfied, particularly when the dimension of $H$ is large with respect to the dimension of $\params$.
Furthermore, downstream observables benefiting from the same affine representation can be quickly emulated via $\braket{\psi | O | \psi} \approx \coeffs^\dagger [X^\dagger O X] \coeffs$ where the factor in brackets is computed and stored in the offline stage.
Thus, the time spent ``training'' the reduced-order model (constructing the basis $X$ and projecting $H$ and $O$) in the offline stage can lead to \emph{multiple} fast \& accurate emulators for the energy and other observables.

\subsection{Results from other fields}
\label{sec:results_eigen}

The Hermitian and symmetric eigenvalue problems have been extensively studied in the mathematical literature and thus the ability to approximate eigenvalues and eigenvectors from a subspace is well known~\cite{parlett1980symmetric,Jia2001AnAO}.
For example, the convergence properties of eigenvalues and eigenvectors, along with their error bounds, are discussed under these subspace projections in the linear algebra and applied mathematics literature.
These analyses extend beyond the extremal eigenvalues, but to all eigenvalues in the spectrum of $H$, where the applicability of subspace approaches to excited states has been clear for decades.
The sense in which the Ritz values and vectors are \emph{optimal} approximations is well known, and, \eg, discussed in Refs.~\cite{parlett1980symmetric,Beattie2000eigenGalerkin}.

Furthermore, eigenvalue problems had been studied in the field of PMOR well before EC was introduced into the nuclear literature.
For example:
\begin{itemize}
\item Machiels~\etal~\cite{MACHIELS2000153} applied a parametric snapshot-based reduced-order approach to quickly evaluate eigenvalue problems, and subsequently proved theorems about the error bounds of the approximate eigenvalues. Horger~\etal~\cite{Horger2017RBMeigenvalue} built upon this approach for error bounds of multiple eigenvalues and discusses efficient greedy algorithms for the basis generation.
\item Pau~\cite{Pau2007RBMBandStructure} used this reduced-basis approach to quickly and accurately compute eigenvalues for band structure calculations.
\item Buchnan~\etal~\cite{Buchan2013EigenvaluePOD} constructed a reduced-order model for parametric eigensystems in reactor physics.
\item Cheng~\cite{Cheng2016SchrodingerRBM} constructed reduced-order models of the bound-state Schr\"odinger equation for electron wave functions in semiconductor nanostructures.
\item Gr\"abner~\etal~\cite{grabner2015} emulated a parameterized non-linear eigenvalue problem using a snapshot-based reduced-order model for resonant frequencies.
\end{itemize}

Eigenvalue problems in PMOR appear during the greedy sampling (see Sec.~\ref{sec:model_reduction}) and uncertainty quantification phases of reduced-order models for partial differential equations~\cite{HUYNH2007473,Rozza2008}.
These phases require, in part, the solution to a generalized eigenvalue problem to obtain the coercivity constant $\alpha(\params)$ relevant for the error bounds~\cite{chen2017RBUQ,Haasdonk2016RBM,HUYNH2007473,Rozza2008}, where $\alpha(\params)$ is the minimum eigenvalue.
A fast approximation for $\alpha(\params)$ is needed across many values of $\params$ and hence calculations in the large space must be avoided.
References~\cite{HUYNH2007473,Rozza2008} considered the reduced-order model [Eq.~\eqref{eq:generalized_eigenvalue}] for $\alpha(\params)$ but ultimately proposed instead the so-called Successive Constraint Method (SCM) due to its ability to provide a rigorous lower bound to $\alpha(\params)$.
Nevertheless, others still encourage the use of Eq.~\eqref{eq:generalized_eigenvalue} due to its speed~\cite{Hess7254247} and simplicity~\cite[Ch.\ 4]{Benner20201} over the SCM.

Each of the examples provided in this section contain all of the key ingredients of EC---creating snapshots at parameter values, projecting large eigensystems to the span of these snapshots, evaluating rapidly in an online phase, etc.---but have been known as the reduced basis method in the model reduction community.

\section{Variational Emulators}
\label{sec:variational}

\subsection{Theory}
\label{sec:variational_theory}

Variational principles are ubiquitous in physics.
Many differential equations have a corresponding action~$\action$, where the solution to the differential equation also makes~$\action$ stationary.
This yields an alternate way of solving a set of PDEs: rather than solving the Euler-Lagrange equations themselves, one can instead find the solution that makes the action stationary under variations in $\psi$.
The use of variational principles as a means to solve otherwise difficult problems dates back to Ritz~\cite{Ritz1909,Ritz1909_2,LEISSA2005961,ILANKO2009731}.
Thus, these methods often go under the names of (Rayleigh-)Ritz~\cite{LEISSA2005961,ILANKO2009731}, or are simply described as variational.
But as we will see in Sec.~\ref{sec:galerkin}, the Galerkin approach is more general and hence these are occasionally named Ritz-Galerkin methods.
Here we provide a brief description of how reduced-order models can arise from variational principles; for an extensive discussion of variational methods, see Ref.~\cite{zienkiewicz2013finite}.

One can derive a set of differential equations---Euler-Lagrange equations---from a variational principle~\eqref{eq:action} by enforcing $\delta\action = 0$ under arbitrary variations $\delta\psi$.
However, such differential equations may require a fine grid or otherwise be expensive to solve.
Instead we would like to obtain a set of reduced-order models directly from the variational principle.
To do so we note that variations $\delta\psi$ can no longer be completely arbitrary given our choice of trial function in Eq.~\eqref{eq:trial_general}.
Rather than stipulate that $\delta\action = 0$ for any arbitrary variation $\delta\psi$, we instead extract the optimal coefficients, $\coeffs_\star$, as those for which $\action$ is stationary under variations in $\coeffs$:
\begin{align} \label{eq:action_stationary_ansatz}
    \delta\action[\trial\psi] = \sum_{i=1}^{\nbasis} \frac{\partial\action}{\partial\coeff_i}\delta\coeff_i = 0.
\end{align}
This yields a set of $\nbasis$ equations and unknowns $\coeffs$ because the factor multiplying each $\delta\coeff_i$ must be identically zero.

The general case would involve a numerical search for the solution to Eq.~\eqref{eq:action_stationary_ansatz}, but if $\action$ is quadratic in $\psi$ then the solution can be determined analytically.
We will focus here on the case of solving a differential equation, which will result in a linear problem to be solved, because we have already tackled the eigenvalue problem in Sec.~\ref{sec:eigen}.
In this case, $\action$ can be written as
\begin{align}
    \action[\trial\psi] & = \frac{1}{2}\braket{\trial\psi | A | \trial\psi} + \braket{b | \trial\psi} + c \\
    & = \frac{1}{2}\coeffs^\dagger \subspace A \coeffs + \vec{b}\cdot \coeffs + c, \label{eq:S_in_reduced_space}
\end{align}
where $\subspace A = X^\dagger A X$, $b_i = \braket{b | \psi_i}$, and $c$ is a constant.
The quadratic portion could be made symmetric---if it is not already---by writing it as
\begin{align}
    \action & = \frac{1}{2}\coeffs^\dagger \subspace A_s \coeffs + \vec{b}\cdot \coeffs + c, \\
    \subspace A_s & = \frac{\subspace A + \subspace A^\dagger}{2},
\end{align}
which can be desirable for numerical purposes.
It then follows that the optimal coefficients $\coeffs_\star$ are those that satisfy
\begin{align} \label{eq:coefficient_solve_quadratic}
    \delta\action = \subspace A_s \coeffs_\star + \vec{b} = 0,
\end{align}
which can be solved with standard linear algebra methods.
This linear equation is of dimension $\nbasis$, the number of basis elements $\{\psi_i\}$, rather than of the much larger dimension of $\psi$ itself.
Therefore, as long as $\{\psi_i\}$ approximately span the space that $\psi$ traces as a function of $\params$, the trial function constructed by Eqs.~\eqref{eq:trial_general} and~\eqref{eq:coefficient_solve_quadratic} will be a fast \& accurate emulator of $\psi$.

This method is particularly beneficial for quickly emulating many $\params$ values if both $A$ and $\ket{b}$ are \emph{affine} in $\params$, that is
\begin{align}
    A(\params) & = \sum_n f_n(\params) A_n, \label{eq:A_affine} \\
    \ket{b(\params)} & = \sum_n g_n(\params) \ket{b_n}, \label{eq:b_affine}
\end{align}
which need not contain the same number of terms and, from which,
\begin{align}
    \subspace A(\params) & = \sum_n f_n(\params) \subspace A_n, \label{eq:A_tilde_def} \\
    \vec{b}(\params) & = \sum_n g_n(\params) \vec{b}_n  
\end{align}
can be quickly reconstructed because $\subspace A_n = X^\dagger A_n X$ and $b_{ni} = \braket{b_n | \psi_i}$ need only be computed once in the offline stage.
This follows similarly from the discussion in Sec.~\ref{sec:eigen}, but in this case both $A$ and $\ket{b}$ can depend on the parameters of the system, rather than simply the Hamiltonian $H$.

Note that the number of basis elements $\nbasis$ needed for an accurate $\trial\psi$ may be much smaller than one might na\"ively expect.
Even if the dimension of a high-fidelity solution $\psi$ is quite large (\eg, due to a fine grid size in the differential equation solver), the space that $\psi$ traces out as a function of $\params$ is often much smaller.
Thus, constructing an emulator for accurately reproducing $\psi(\params)$ can be achievable with a well-chosen basis $X$ (see Sec.~\ref{sec:model_reduction}).

\subsection{Constraints}
\label{sec:variational_constraints}

The derivation in Sec.~\ref{sec:variational_theory} assumed that $\psi$ was unconstrained, but oftentimes one has to enforce a set of constraints $C_j(\psi)=0$ for $j = 1,\cdots, N_c$.
For example, in the eigenvalue problem example in Sec.~\ref{sec:eigen}, we had to enforce that the wave function is normalized to one, \ie, $C(\psi) = \braket{\psi | \psi} - 1 = 0$.
Constraints can be straightforwardly included in a variational principle via the method of Lagrange multipliers.
Here, each constraint is appended as a term in the variational form with a corresponding $\lambda_j$, \ie, $\lambda_j C_j(\psi)$.
When imposing stationarity, each of these terms yields a $\delta\lambda_j C_j(\psi) + \lambda_j \delta C_j(\psi)$ contribution to $\delta\action$.
Specifically, Eq.~\eqref{eq:action_stationary_ansatz} needs to be rewritten as
\begin{align} \label{eq:action_stationary_ansatz_lagrange}
    \delta\action[\trial\psi] = \sum_{i=1}^{\nbasis} \frac{\partial\action}{\partial\coeff_i}\delta\coeff_i + \sum_{j=1}^{N_c} \frac{\partial\action}{\partial\lambda_j}\delta\lambda_j = 0,
\end{align}
from which follows a set of $\nbasis + N_c$ equations.
Enforcing these constraints then yields a larger system of equations to solve when emulating $\psi$ in the online phase, though in some systems it is possible to solve for $\lambda_j$ in terms of $\psi$, hence reducing the problem back to its original size while still incorporating the constraints~\cite{zienkiewicz2013finite}.
Solving for $\lambda_j$ to remove it from $\action$ is beneficial if possible because (i)~it will decrease the size of the linear system to be solved and (ii)~it can make the system better conditioned numerically.

\subsection{Concrete Example}
\label{sec:variational_concrete_example}

At this point it is helpful to provide a concrete example of a variational principle that leads to a reduced-order model.
Here we provide a simple projection example without complicating details or abstract notation; see Ref.~\cite{zienkiewicz2013finite} for more examples.

Consider the functional, given functions $g$ and $f$~\cite{jackson_classical_1999},
\begin{align}
    \action = \int_\Omega \dd{\Omega} \left[\frac{1}{2} \nabla\psi \cdot \nabla\psi - g\psi\right] - \int_{\Gamma} \dd{\Gamma} f\psi.
\end{align}
Under an infinitesimal variation $\delta\psi$, the change $\delta\action$ is then
\begin{align}
    \!\!\delta\action & = \int_\Omega \dd{\Omega} [\nabla\delta\psi \cdot \nabla\psi - g\delta\psi] - \int_{\Gamma} \dd{\Gamma} f\delta\psi\\
    & = \int_\Omega \dd{\Omega} \delta\psi\left[ -\nabla^2\psi - g\right] + \int_{\Gamma} \dd{\Gamma} \delta\psi \!\left[\frac{\partial\psi}{\partial n} - f\right],
\end{align}
from which it follows that the stationary solution is
\begin{align}
    -\nabla^2\psi = g \quad \text{in } \Omega, \label{eq:poisson_domain} \\
    \frac{\partial\psi}{\partial n} = f \quad \text{on }\Gamma, \label{eq:poisson_boundary}
\end{align}
which is exactly the Poisson equation with Neumann boundary conditions.

By instead starting with $\action[\trial\psi]$ and imposing Eq.~\eqref{eq:action_stationary_ansatz}, then it follows that
\begin{align} \label{eq:poisson_action_reduced_variational}
    \delta\action = \delta\coeff_i \Big[\int_\Omega \dd{\Omega} \left[ (\nabla \psi_i) \cdot (\nabla \psi_j)\coeff_j - g\psi_i\right] - \int_{\Gamma} \dd{\Gamma} f\psi_i \Big].
\end{align}
With $\subspace A_{ij} \equiv \int_\Omega\dd{\Omega} (\nabla \psi_i) \cdot (\nabla \psi_j)$, $g_i \equiv \int_\Omega \dd{\Omega} g \psi_i$, and $f_i \equiv \int_\Gamma \dd{\Gamma} f \psi_i$ the reduced-order model becomes
\begin{align} \label{eq:poisson_reduced_basis_variational}
    \subspace A\coeffs_\star = \vec{g} + \vec{f},
\end{align}
which is an explicit example of Eq.~\eqref{eq:coefficient_solve_quadratic}.
This could provide a fast \& accurate emulator of $\psi(\params) \approx X\coeffs_\star(\params)$ for systems where $g$ or $f$ are affine functions of the parameters $\params$.

\subsection{Results from other fields}
\label{sec:variational_results_other_fields}

Many examples from Secs.~\ref{sec:results_eigen} and~\ref{sec:galerkin_results_other_fields} could be listed here due to the relationship between emulators constructed from variational principles and Galerkin methods.
Instead we highlight an explicitly variational problem particularly relevant for nuclear systems: density functional theory (DFT).

Computing the ground state energies in systems of nuclei and/or electrons reduces again to a minimization problem, where the wave functions---or equivalently densities---are those that minimize the non-linear functional $\mathcal{E}$.
These minimization problems appear in quantum chemistry and nuclear physics and are often approached in the DFT framework~\cite{engel2011density,10.1088/2053-2563/aae0ed}.
The emulation of DFTs via reduced-order models has been studied in the quantum chemistry literature, \eg, in Refs.~\cite{WOS:000208297800036,cances2007feasibility} (see also Refs.~\cite{Pau2007RBMBandStructure,LIN20122140,ZHANG2017426}), where an empirical interpolation method (see Sec.~\ref{sec:model_reduction}) was used to avoid issues with the nonlinearities of $\mathcal{E}$.
This interpolation method permitted the reduced-order model to project all large-space operators to the reduced space of snapshots up front, which is a critical step to retain an efficient online-offline decomposition in the emulator.

\subsection{Results from nuclear physics}
\label{sec:variational_results_nuclear}

We have already discussed how the eigenvalue problem can be projected to a subspace by starting with a variational principle in Sec.~\ref{sec:eigen}, which corresponds to computing the energy spectrum and wave functions in a bound nuclear system.
Its suitability for uncertainty quantification in low-energy nuclear physics has been demonstrated, \eg, in chiral EFT applications to few- and many-body systems, where it has been described as EC~\cite{Konig:2019adq, Ekstrom:2019lss, Wesolowski:2021cni}.
Variational approaches for reduced-order models~\cite{Furnstahl:2020abp,Drischler:2021qoy,Zhang:2021jmi} have also been very successfully applied to \emph{scattering} states via the Kohn variational principle~\cite{taylor2006scattering}, where the Schr\"odinger equation is no longer an eigenvalue problem (for a brief review see Ref.~\cite{Drischler:2022yfb}).
Furthermore, a new approach for emulating directly scattering $K$ or $T$ matrices (without trial wave functions) was proposed in Ref.~\cite{Melendez:2021lyq} based on the Newton variational principle (NVP)~\cite{newton2002scattering}.
Each of these methods has benefited from an affine parameterization of the Hamiltonian $H(\params)$, which permitted fast \& accurate emulation of nuclear observables.

In addition, the emulation of scattering states with non-affine parameterization was also studied in Ref.~\cite{Zhang:2021jmi}, where $\subspace A (\params)$ (defined in Eq.~\eqref{eq:S_in_reduced_space}) does not satisfy Eq.~\eqref{eq:A_tilde_def}. In order to reduce the computing costs in the online phase, a Gaussian process (GP)~\cite{rasmussen2006gaussian} was used to emulate $\subspace A (\params)$ across the parameter space.
The training of the GP increased the offline computing costs, but the online costs were similar to those with affine parameterizations.
Since the $\params$ dependence for $\subspace A (\params)$ is much smoother than those for $\psi(\params)$ and $\subspace \psi(\params)$, the number of GP training points are kept to a minimum while still achieving great accuracy.

To our knowledge, the first application of projection-based PMOR as applied to nuclear DFT was presented in Ref.~\cite{Melendez_Quantum_Emulator_Examples_2021} (see Ref.~\cite{Melendez_Reduced-Order_DFT_Emulators_2021} for the code).
Here, a trial density was proposed as a linear combination of exact densities $\{\rho_i\}$ at a set of training locations $\params_i$.
The coefficients $\coeffs$ were found as those that minimized the energy of the system given a set of parameters at which to emulate.
Because this is a non-quadratic variational problem, such coefficients were found empirically via an optimizer.
The goal of this work was to use a proof of principle to advocate for the adoption of such projection-based tools by the nuclear community outside of bound-state and scattering systems connected to Hamiltonians.

\section{Galerkin Emulators}
\label{sec:galerkin}

\subsection{Theory}

The Galerkin approach, also more broadly called the ``method of weighted residuals,'' relies on the \emph{weak} formulation of the differential equations in Eq.~\eqref{eq:generic_differential_and_boundary} rather than a variational principle.
To obtain the weak form, the differential equation and boundary condition are multiplied by arbitrary test functions $\phi$ and $\bar\phi$, integrated over the domain and boundary, and their sum set equal to zero:
\begin{align} \label{eq:weak_differential}
    \int_\Omega \dd{\Omega} \phi  D(\psi) + \int_\Gamma \dd{\Gamma} \bar\phi  B(\psi) = 0.
\end{align}
If Eq.~\eqref{eq:weak_differential} holds for all $\phi$ and $\bar\phi$, then Eq.~\eqref{eq:generic_differential_and_boundary} must be satisfied as well.
The form of Eq.~\eqref{eq:weak_differential} is often rewritten using integration by parts to reduce the order of derivatives and to simplify the solution.
Importantly, the weak form has the integral form needed for our emulator application.
The weak form and its Galerkin projection are used extensively, for example, in the finite element method; see Refs.~\cite{zienkiewicz2013finite,Zienkiewicz2014finitesolid,Zienkiewicz2014finitefluid} for an in-depth study and list of examples.
Here we follow the introduction of Galerkin methods as provided in Ref.~\cite{zienkiewicz2013finite}.

Starting with the weak form, we can begin to construct an emulator that avoids the need for an explicit variational principle.
It begins by first noting that substituting our trial function~\eqref{eq:trial_general} into $D(\psi)$ and $B(\psi)$ will not in general satisfy Eq.~\eqref{eq:generic_differential_and_boundary} regardless of the choice of $\coeffs$.
Therefore, there will be some \emph{residual}, and the goal is to find $\coeffs_\star$ which minimize that residual across a range of test functions $\phi$ and $\bar\phi$.
This system would be over-determined in the case of truly arbitrary test functions, so instead we propose the test bases
\begin{align}
    \phi & = \sum_{i=1}^{\nbasis} \delta\coeff_i\phi_i,
    \qquad
    \bar\phi = \sum_{i=1}^{\nbasis} \delta\coeff_i\bar\phi_i,
\end{align}
where $\delta\coeff_i$ are arbitrary parameters, not related to $\coeff_i$.
The $\delta\coeff_i$ will play the same role as those in Eq.~\eqref{eq:action_stationary_ansatz}, namely as a bookkeeping method for determining the set of equations that are equivalently zero.
By enforcing that the residuals against these test functions vanish for arbitrary $\delta\coeff_i$, the bracketed expression in 
\begin{align} \label{eq:weak_form_subspace}
    \delta\coeff_i \Bigl[\int_{\Omega} \dd{\Omega}  \phi_i  D(X\coeffs_\star) +  \int_{\Gamma} \dd{\Gamma} \bar\phi_i  B(X\coeffs_\star)
    \Bigl]= 0,
\end{align}
is zero for all $i \in [1, \nbasis]$, from which the optimal $\coeffs_\star$ are extracted.
Because this approximately satisfies the weak formulation, we have found an approximate solution to Eq.~\eqref{eq:generic_differential_and_boundary}.

In a variety of cases~\cite{zienkiewicz2013finite}, the test function basis is chosen to coincide with the trial function basis $X$, \ie, $\phi_i = \bar\phi_i = \psi_i$.
This particular choice is known as \emph{the} Galerkin method, but it is sometimes further specified as the Ritz-Galerkin or Bubnov-Galerkin methods.
However, the method of weighted residuals is more general than the variational methods described in Sec.~\ref{sec:variational} because the test space need not be equivalent to the trial space (\ie, $\phi_i \neq \psi_i$).
In these cases, the approach is described as the Petrov-Galerkin method~\cite{zienkiewicz2013finite};
this can result in more efficient emulators for some differential equations~\cite{Zienkiewicz2014finitefluid}.

Under the Ritz-Galerkin assumption for the test space we can derive the reduced-order model for the case of a linear operator: $D(\psi) = D\ket{\psi} + \ket{b}$.
If we ignore the boundary condition for simplicity, it then follows from Eq.~\eqref{eq:weak_form_subspace} that
\begin{align}
    \subspace D \coeffs_\star + \vec{b} = 0,
\end{align}
where $\subspace D = X^\dagger D X$ and $b_i = \braket{b | \psi_i}$.
Just like in Sec.~\ref{sec:variational}, we have arrived at a linear problem for the solution to $\coeffs_\star$ and insofar as $\nbasis$ is small compared to the size of $\psi$, this will yield improvements to the time it takes to obtain a solution for $\coeffs_\star$.
Further speedups are available if $D$ and $\ket{b}$ are affine in the parameters $\params$ so that $\subspace D$ and $\vec{b}$ can be efficiently recomputed in the online phase---see Secs.~\ref{sec:eigen} and~\ref{sec:variational}.

\subsection{Concrete Example}
\label{sec:galerkin_concrete_examples}

Here we repeat the example provided in Sec.~\ref{sec:variational_concrete_example} but instead start from the Poisson equation [Eqs.~\eqref{eq:poisson_domain} and~\eqref{eq:poisson_boundary}] and then derive the weak form.

First we multiply each equation by a test function $\phi=\bar\phi$, integrate over the respective domains, and add the equations together:
\begin{align}
    \int_\Omega \dd{\Omega} \phi \left[-\nabla^2\psi - g\right] + \int_\Gamma \dd{\Gamma} \phi \left[\frac{\partial\psi}{\partial n} - f\right] = 0.
\end{align}
Next we use the divergence theorem to symmetrize the system and to reduce the order of the derivatives:
\begin{align} \label{eq:weak_poisson}
    \int_\Omega \dd{\Omega} \left[\nabla \phi \cdot \nabla\psi - g\phi\right] - \int_\Gamma \dd{\Gamma} f\phi = 0,
\end{align}
which is the weak form we desire.
Finally, by asserting that Eq.~\eqref{eq:weak_poisson} holds for $\psi \to \trial\psi = X\coeffs$ and $\phi = \sum_i \delta\coeff_i \psi_i$ for $i \in [1,\nbasis]$, then we have its discretized form
%\begin{align}
    % \int_\Omega \dd{\Omega} \delta\coeff_i\left[\nabla\psi_i \cdot \nabla \psi_j \coeff_j - g\psi_i\right] - \int_\Gamma \dd{\Gamma} \delta\coeff_i f \psi_i = 0,
%\end{align}
\begin{align}
    \delta\coeff_i\Bigl[
    \int_\Omega \dd{\Omega} \left[\nabla\psi_i \cdot \nabla \psi_j \coeff_j - g\psi_i\right] - \int_\Gamma \dd{\Gamma}  f \psi_i 
    \Bigr] = 0,
\end{align}
which is exactly Eq.~\eqref{eq:poisson_action_reduced_variational} and~\eqref{eq:poisson_reduced_basis_variational} found via the variational approach!

\subsection{When Galerkin Coincides with Variational Emulators}
\label{sec:equivalence}

We have already seen that the eigen-emulators of Sec.~\ref{sec:eigen} could be derived by both variational and Galerkin procedures, and we have found that the exact same reduced-order model for the Poisson equation arises in both the variational and the Galerkin procedures.
This raises the question: when is the variational approach equivalent to the Galerkin approach?

The answer becomes clear if one restricts to cases where Eq.~\eqref{eq:generic_differential_and_boundary} is the Euler-Lagrange equation derived from the action~\eqref{eq:action}.
By the definition of the Euler-Lagrange equations, this statement is equivalent to
\begin{align} \label{eq:action_stationary_full}
    \delta\action[\psi] = 0 = \int_\Omega \dd{\Omega} \delta\psi D(\psi) + \int_\Gamma \dd{\Gamma} \delta\psi B(\psi).
\end{align}
We can then consider changing variables from the trial stationarity condition~\eqref{eq:action_stationary_ansatz}
\begin{align}
    \delta\action[\trial\psi] = 0
    = \sum_{i=1}^{\nbasis}\frac{\partial\action}{\partial\coeff_i}\delta\coeff_i
    = \sum_{i=1}^{\nbasis}\left[\frac{\partial\action}{\partial\psi}\right]_{\psi=\trial\psi} \frac{\partial\trial\psi}{\partial\coeff_i}\delta\coeff_i,
\end{align}
where the first two factors on the right-hand side are Eq.~\eqref{eq:action_stationary_full} with $\delta\psi \to \partial\trial\psi/\partial\coeff_i = \psi_i$ and thus
\begin{align}
    0 = \delta\coeff_i\int_\Omega \dd{\Omega} \psi_i D(X\coeffs_\star) + \delta\coeff_i\int_\Gamma \dd{\Gamma} \psi_i B(X\coeffs_\star),
\end{align}
for all $i \in [1, \nbasis]$.
This is exactly Eq.~\eqref{eq:weak_form_subspace} under the Galerkin assumption that $\phi_i = \bar\phi_i = \psi_i$.
Note that $\action$ was not assumed to have a quadratic form for this derivation, and no linearity assumptions were made about $D$ or $B$.

There are some final steps to proving that Galerkin is strictly more general than a variational (Ritz) approach.
The first is that all variational principles have corresponding Euler-Lagrange equations.
This can be shown via the fundamental lemma of the calculus of variations.
Second, there is the question of how Lagrange multipliers are to be introduced in the weak form without first starting with a variational principle.
In this case, one simply adds the relevant constraint terms to the weak form.
That is, if the constraint $C(\psi)=0$ would appear as $\lambda C(\psi)$ in the functional $\action$, then one would simply add $\delta\lambda C(\psi) + \lambda \delta C(\psi)$ to the weak form and use the fact that $\delta\trial\psi = X\delta\coeffs$.
Therefore, if the differential equations to be solved directly correspond to the Euler-Lagrange equations of an action, then the variational approach is identical to the Galerkin approach.
The generality of the Galerkin method comes from the fact that one need not assert that $\phi_i = \bar\phi_i = \psi_i$.

Although we have shown that the Galerkin method is more general than the Ritz approach, there is still value to obtaining a variational principle and deriving a reduced-order model from it.
First, the functional $\action$ is often physically meaningful in its own right, and hence the variational emulators provide a straightforward way of quickly computing its value.
For example, in the NVP approach to emulating the scattering $K$ (or $T$) matrix~\cite{Melendez:2021lyq}, both the trial function $\trial K$ and the NVP functional $\mathcal{K}[\trial K]$ at the stationary point are estimates of $K$, but the variational principle has better error properties: if $\trial K$ has an error of $\mathcal{O}(\delta K)$ then $\mathcal{K}$ has an error of order $\mathcal{O}(\delta K^2)$.
Thus, obtaining and using the variational principle as an emulator for $K$ is superior to simply applying the Galerkin method.
Second, the variational emulators are guaranteed to provide symmetric matrices when solving for $\coeffs$.
This feature can provide numerical benefits both when constructing the relevant matrices and when solving for $\coeffs$.

\subsection{Results from other fields}
\label{sec:galerkin_results_other_fields}

We have found that examples of reduced-order models relying on a Galerkin projection are numerous and exist across a multitude of disciplines.
A wide array of such examples are given in Ref.~\cite{Benner2020Volume3Applications}, which is an entire volume of a 3-part series dedicated to the applications of MOR in diverse settings, ranging from engineering to life sciences.
A comprehensive list of reduced-order models built from Galerkin methods would be impossible; we instead provide a curated list of helpful articles below.

A broad survey on parametric model reduction~\cite{benner2015survey} cites multiple examples of highly successful applications of reduced-order models, some of which we highlight here.
In the thermal modeling of electric motors that depends on 20 parameters, Bruns~\etal~\cite{Bruns2015PROMthermal} use model reduction for a speedup over 300--500 times the high-fidelity model.
Lassila~\etal~\cite{Lassila2013reducedHemo} display the power of reduced-order modeling in the study of nonlinear viscous flows by reducing high-fidelity models of dimension $>10,000$ to emulators of dimensions 8--20, with minimal accuracy impact and speedup factors of up to 450.
Convection-diffusion models arise in the investigation of contaminant transport, where Lieberman~\etal~\cite{Lieberman2013reducedContaminant} exploit MOR to reduce the model order from over one million down to 800 with negligible accuracy loss and a 3,000 speedup factor.

Rozza~\etal~\cite{Rozza2008} provide an illuminating introduction and motivation to reduced-order models built from Galerkin projections.
Its applications include heat conduction and convection-diffusion, inviscid flow, and linear elasticity systems.
Beyond the specific applications listed here, Ref.~\cite{Rozza2008} describes how to effectively choose parameter values $\params$ for building the basis via a greedy algorithm, explores issues of convergence and error bounds, and performs an analysis of computational costs.

Chen~\etal~\cite{chen2017RBUQ} review the state of the reduced basis method literature, collecting many of the main ingredients necessary for building and analyzing reduced-order models.
They demonstrate their claims via benchmark problems and describe generalizations to time-dependent systems, risk prediction, Bayesian inverse problems, and more.
The topic of uncertainty quantification is addressed in detail, which, for example, allows for effective placement of snapshot locations $\psi(\params_i)$ via a greedy algorithm.

\section{The Model Reduction Framework}
\label{sec:model_reduction}

Projection-based PMOR consists of
(1)~sampling across parameters for snapshot candidates, (2)~creating the snapshots and generating a basis $X$, and
%(1)~sampling across parameters $\params$ to create snapshots, (2)~generating a basis $X$ from the snapshots, and
(3)~using the basis to construct the reduced system~\cite{benner2015survey}.
Sections~\ref{sec:variational} and~\ref{sec:galerkin} partially address step (3) by providing two closely related methods of constructing these projected systems, one based on variational principles and another based on the Galerkin method.
Here we take a bird's-eye view of the parametric model reduction workflow to show where different assumptions can lead to different types of reduced-order models, and to point to extensions at the cutting edge of the model reduction literature.
For in-depth reviews, see, \eg, Refs.~\cite{benner2015survey,Benner2017modelRedApprox,Benner2020Volume1DataDriven,Benner20201,Benner2020Volume3Applications}.

\subsection{Sampling Parameters}

Many reduced-order emulators rely on the concept of snapshots, or exact solutions at particular values across a range of parameters $\params$.
So then the question becomes how to wisely choose the snapshot locations $\{\params_i\}$.
For parameter spaces that are not too large, one could employ a space-filling design (such as a latin-hypercube or grid-based approach), or one could center the design close to the range of parameter values that will be emulated.
Define the finite number of points sampled from the space as $\ntrain$.
Ultimately, we will want a set of $\nbasis \leq \ntrain$ snapshots to construct the reduced basis, but the manner in which the snapshots are generated from the $\ntrain$ training parameters differs based on the basis construction method.

\subsection{Constructing a basis}

The next step is to take the snapshots and to construct a basis $X$ given the set of training parameters.
Notably, we have mostly restricted our attention to systems without time dependence.
In the static, time-independent case, the sampling need only occur in the space of parameters $\params$.
Here we discuss two approaches for determining the ``optimal'' $\nbasis$ basis vectors: Proper Orthogonal Decomposition (POD) and greedy algorithms.

POD (\aka principal component analysis)~\cite{Gubisch2017} is an explore-and-compress strategy used to extract the most important basis vectors from a set of snapshots~\cite{hesthaven2015certified}.
It computes snapshots $\psi(\params_i)$ at \emph{all} $\ntrain$ parameter values and subsequently keeps only the $\nbasis$ most important vectors.
It performs a singular value decomposition of the $\ntrain$ snapshot vectors and then removes those least important to the spanning set, \ie, those with the smallest eigenvalues.
Often the set of important vectors are chosen such that the percent of the remaining ``energy'' (sum of the eigenvalues) is large, say, 99\% relative to the total set of vectors.
The orthonormalization performed during the POD step is also helpful during the online phase of the emulator because it can improve the conditioning of the system.
Because POD evaluates the snapshots at all of the proposed $\ntrain$ parameter values before then compressing the information, this can either be wasteful of computing resources or severely limit the size of $\ntrain$.
Hence, we now turn to an alternative.

The greedy algorithm for basis generation is an iterative approach~\cite{Rozza2008,hesthaven2015certified,chen2017RBUQ} that does not evaluate the high-fidelity model at all $\ntrain$ parameter values.
At each step the next location $\params_i$ to take a snapshot is chosen to be that which is expected to minimize the error in the emulator.
Critical to such an approach is a fast approximation to the emulator error.
Uncertainty quantification for reduced-order models have been well studied and, \eg, are available for parabolic and elliptic PDEs~\cite{chen2017RBUQ,hesthaven2015certified}, and the eigenvalue problem~\cite{Horger2017RBMeigenvalue,sarkar2021selflearning}.
At each step, the error at the set of $\ntrain$ parameter values is estimated, and that with the largest expected error is then evaluated with the high-fidelity model and appended to the basis.
The search stops once the desired error tolerance has been achieved, or after a given number of steps.
Because the error estimate is built to be much faster than the high-fidelity model, this approach is often much more efficient than the POD method.

Now we move on to the time-dependent cases, where sampling can occur not only in parameter space, but also in time (or frequency).
In the time-dependent case there are more options, such as the rational interpolation method~\cite{Baur2011projectInterpolate,Baur2017TimeDep}, whose snapshots will include samples from the frequency domain and the parameter domain.
Additionally, there exist POD variants for both the time and the frequency domains~\cite{benner2015survey}.
Note that mixed approaches exist: for example, in a POD-greedy approach one can opt to use a greedy algorithm in an ``outer loop'' in parameter search, while using a POD-based approach to evaluate each time snapshot before discarding the least important~\cite{hesthaven2015certified}.
Another common technique is known as balanced truncation, which creates a set of reduced models at each parameter value, which are then interpolated to create an emulator across $\params$~\cite{benner2015survey}.

\subsection{Building the reduced-order model}
\label{sec:Building_the_ROM}

Finally, the reduced model must be created from the basis.
In the examples shown above,  where the snapshots are collected into a single basis $X$ and the operators are projected to this basis, this is straightforward.
This can work well if the system is linear and the operators are affine with respect to the parameters $\params$, which critically permits the reduced-order model to be independent of the size of the high-fidelity space during the online phase.
However these conditions are not always satisfied.

In the non-affine or non-linear cases, one can turn to so-called \emph{hyperreduction} approaches to construct approximate affine representations, which trade accuracy for speed~\cite{Quarteroni:218966}.
These can be classified into the ``approximate then project'' or the ``project then approximate'' classes.
The approximate-then-project approach first approximates the non-affine or non-linear operators as a linear combination of affine operators whose coefficients are to be determined, and whose operators can then be projected via $X$ into the small space up front.
Some common approximate-then-project methods~\cite{hesthaven2015certified} include the empirical interpolation method (EIM)~\cite{Barrault2004,Grepl2007}, the discrete EIM (DEIM)~\cite{Chaturantabut2009,Chaturantabut2010}, and the gappy-POD method~\cite{gappyPOD2003,Carlberg2011gappyPOD}.
Project-then-approximate approaches have been developed more recently and attempt to interpolate the basis $X$ or the projected operators $\subspace{H}(\params) = X^\dagger H(\params) X$ themselves; see Refs.~\cite{Amsallem2010,AnCubature2008,Farhat2014,Benner20201} for examples.
Outside of explicit projection-based approaches, non-intrusive methods have been proposed for dealing with non-linear and non-affine systems, including the use of machine learning tools to approximate the basis coefficients $\coeffs$ or projected operators $\subspace{H}$, \eg,~see Refs.~\cite{GUO2018807,Zhang:2021jmi}.

But in fact, one need not create one basis and one reduced system; rather one could partition $X$ into multiple bases across the span of $\params$.
This can help alleviate the curse of dimensionality in the parameters $\params$ and the computational costs during the offline stage.
Additionally, it permits the use of variable fidelity bases in different parts of the domain.
If one opts to create a set of local systems, each component must be then coupled with one another across the interfaces.
For a survey on this topic, see Ref.~\cite{Benner20201}.

\subsection{Discussion}

Now that we have introduced many of the common steps in model reduction, we can begin to contextualize other named methods, such as the Reduced Basis (RB) method and EC\@.
The RB method was initially introduced in Ref.~\cite{Noor1980reducedBasisFirst} and has found widespread use in the emulation of PDEs in a reduced-order approach~\cite{hesthaven2015certified}.
Due to its similarities with many of the methods discussed here and in the model reduction literature, it may be difficult to distinguish between the RB method and more general model reduction techniques.

In fact, the RB method corresponds to specific choices in the model reduction framework~\cite{benner2015survey}.
First, the parameter set for the RB method is often chosen using a greedy algorithm with the help of a fast error estimate, though POD approaches are sometimes adopted, particularly in time-dependent systems where a POD-greedy combination is employed~\cite{hesthaven2015certified}.
Next, a single basis $X$ is constructed out of snapshots and often orthonormalized for stability.
Finally, the RB model is built from a global basis projection, \ie, the same basis is used for the entire space of $\params$.
These are but one of many choices that can be made at each step in the construction of a reduced-order model.

Likewise, we are able to help place EC into its proper context.
EC is a parametric reduced-order model for an eigenvalue problem.
It uses a global basis that is constructed with a snapshot-based POD approach.
The ``active learning'' approach proposed in Ref.~\cite{sarkar2021selflearning} is the inclusion of a greedy sampling algorithm to obtain the next parameter value $\params_i$.
Each of these are common throughout the model reduction literature and have been studied for eigenvalue problems; therefore EC is a specific implementation of the RB method to construct a reduced-order model.

We conclude by noting that reduced-order emulators have pros and cons.
First, these emulators work better in some systems than others---where the quantity of interest lies in on a low-dimensional manifold compared to the size of $\psi$, and where operations on the large space of $\psi$ can be avoided during the online phase.
As one might expect, the relative size of the low-dimensional representation depends on the specific differential equation.
For example, highly non-linear equations may not permit the same low-dimensional representation.
Further, how the operators depend on the parameters could be critical to the effectiveness of the reduced system in two distinct ways: (i) the relative smoothness of the parameter dependence impacts the ability of $\psi$ to live in a low-dimensional representation, and (ii) affine parameter dependence, or at least an effective hyperreduction approach, is critical to avoiding matrix multiplication in the large space of $\psi$.
Lastly, one should not overlook how the quantity of interest itself affects the quality of the low-dimensional representation.
For example, $E$ has better convergence properties than $\ket{\psi}$ in the eigenvalue problem; the RB literature discusses the improvements of focusing on \emph{compliant} quantities of interest~\cite{chen2017RBUQ,Benner20201}.

\begin{table*}[]
\caption{A sampling of recent
%, maintained 
MOR software libraries; see Ref.~\cite[Sec.~13.3]{Benner2020Volume3Applications} for an extensive listing. 
%Mention tutorials as well as demos.
}
\label{tab:software}
\renewcommand{\arraystretch}{1.7}
\begin{ruledtabular}
\begin{tabular}{lllp{5.5cm}}
Library   & Language  & Website & Remark \\
\colrule
pyMOR\footnote{See also the website of the Model Reduction for Parametrized Systems (MoRePaS) collaboration: \href{https://www.morepas.org/}{morepas.org}.}~\cite{milk2016pyMOR}           &    Python       &    \href{https://pymor.org/}{pymor.org}     &   focuses on RBMs for parameterized PDEs; integrates with external PDE solvers  \\
libROM &   \CC       &      \href{http://librom.net/}{librom.net}           & library for efficient MOR techniques and physics-constrained data-driven methods; includes POD, DMD, projection-based ROM, hyper-reduction, greedy algorithm     \\
MORLAB~\cite{BenW21c} & MATLAB & \href{https://www.mpi-magdeburg.mpg.de/projects/morlab}{mpi-magdeburg.mpg.de/projects/morlab} & MOR of dynamical systems based on the solution of matrix equations using spectral projection methods \\
modred~\cite{10.1145/2616912} &   Python       &    \href{https://modred.readthedocs.io/}{modred.readthedocs.io}             &   library for computing modal decompositions and ROMs, including POD, DMD, and Petrov-Galerkin projection     \\
pyROM~\cite{PUZYREV2019157} & Python &  \href{https://github.com/CurtinIC/pyROM}{github.com/CurtinIC/pyROM} &  framework that employs Python visualisation tools; includes POD and DMD\\
pressio~\cite{rizzi2021pressio} & \CC & \href{https://pressio.github.io/}{pressio.github.io} & minimally-intrusive interface for MOR routines, including Galerkin projections %\\
%emgr~\cite{emgr} &  MATLAB &  \href{http://gramian.de}{gramian.de}  &  \\
%KerMor &  MATLAB & \href{https://www.morepas.org/software/kermor/}{morepas.org/software/kermor} & \\
%JaRMoS & Java & \href{https://www.morepas.org/software/jarmos/}{morepas.org/software/jarmos} & \\
%Dune-RB & \CC & \href{http://users.dune-project.org/projects/dune-rb/wiki}{users.dune-project.org/projects/dune-rb/wiki}
\end{tabular}
\end{ruledtabular}
\end{table*}

Beyond the considerations of emulator quality, building an intrusive emulator for complex systems can be a challenge.
There has been great progress in building general software tools for practitioners in reduced-order modeling~\cite{Haasdonk2016RBM,Benner2020Volume3Applications}.
Table~\ref{tab:software} provides a sampling of recent MOR software libraries.
Continuing to develop and publicize such tools will permit greater acceptance of these powerful methods in nuclear physics.

\section{Outlook}
\label{sec:outlook}

The present work provides for nuclear physicists accessible pointers to some of the
relevant literature on projection-based model reduction and shows that solving the parametric eigenvector problem from a subspace is merely a special case of a much broader set of tools.
By properly contextualizing the methods under the projection-based model reduction umbrella, we have shown
how reduced-order models built from a variational principle relate to an equally vast literature on the Galerkin method, which is even more general.
We have not proposed anything novel; 
rather our message is that much information on emulators, some partially rediscovered and more not-yet-applied, can be at our fingertips if we look more widely.

We have shown that the ``reduced basis method'' is the established name of the methods described in the nuclear physics literature as EC, and suggest its adoption.
We believe that using a unified naming convention will not only alleviate confusion due to a conflict of terms used in other fields,
but will permit access to a much broader literature.
It would surely accelerate progress in the application of emulators in the nuclear community~\cite{Bonilla:2022} and facilitate fruitful external collaborations.

\begin{acknowledgments}
We thank Patrick Millican and Daniel Phillips for fruitful discussions and comments on the manuscript.
This work was supported in part by the National Science Foundation under Grant No.~PHY--1913069 and the NSF CSSI program under award
number OAC-2004601 (BAND Collaboration~\cite{BAND_Framework}), and the NUCLEI SciDAC Collaboration under U.S. Department of Energy MSU subcontract RC107839-OSU\@.
This material is based upon work supported by the U.S. Department of Energy, Office of Science, Office of Nuclear Physics, under the FRIB Theory Alliance award DE-SC0013617. 
\end{acknowledgments}

\bibliography{emulator_refs,bayesian_refs}
\end{document}